\documentstyle{article}
\title{Absence of a charge renormalization in the Higgs model
interacting with conformal two-dimensional gravity}
\author{ Z. Haba\\Institute of Theoretical Physics, University of Wroclaw,
\\50-204 Wroclaw, Plac Maxa Borna 9,Poland\\e-mail:zhab@ift.uni.wroc.pl}
\date{}
\begin{document}
\maketitle
\begin{abstract}
We discuss $D$-dimensional scalar and electromagnetic fields
interacting with  a quantized metric.
 The gravitons depend solely on twodimensional
coordinates. We consider a conformal field
theory  as a model for the metric tensor.We show that
 an interaction with gravity
improves the short distance behaviour.
As a result there is no charge renormalization
in the fourdimensional Higgs model.
\end{abstract}
\section{Introduction}
We discuss $D$-dimensional scalar and electromagnetic fields
interacting with  a quantized scale invariant metric.
 The gravitons depend only on twodimensional
coordinates. These coordinates could be considered either as
additional coordinates (e.g., in the brane picture when
gravitons can escape to
extra dimensions \cite{dvali})  or as a (stringy) perturbation of
the physical fourdimensional space-time deforming the flat
Minkowski metric.

In our earlier paper \cite{PLB}
 we have  shown that  as a result of  an interaction
 with a  scale invariant quantum gravity
 the propagators of quantum  matter fields become more
 regular than the ones in a classical  gravitational field.
 In this paper we consider a specific
 model of a scale invariant quantum metric.
 We show
 that as a  result of  an interaction with a
 conformal invariant twodimensional gravity the
 fourdimensional Higgs model has no charge renormalization.
 This property can still hold true in five dimensions
 depending on the scale dimension of the twodimensional
 gravity. Higher dimensional
 models can still be renormalizable
 if we can treat properly
 more singular gravitational fields.
  We also show that after
 an interaction  with twodimensional gravity the standard
 dispersion relation (relating the temporal component of
 the wave  vector to the spatial one) will be changed
 (for some other suggestions see ref.\cite{am}) .
\section{A model of quantum gravity}
We consider a metric tensor on the Riemannian manifold (Euclidean
formulation)
\begin{equation}
(G)^{AB}=g^{AB}
\end{equation}
as a twodimensional field $G$ with values in a set of real symmetric
positive definite $D\times D$ matrices $G$. We choose the metric
in a block diagonal form $G^{AB}=\delta^{AB}$ if $A,B> D-2$ and
for $A,B\leq D-2$ the tensor $ g^{\mu\nu}({\bf x}_{F})$ is a
$(D-2)\times (D-2)$ matrix depending on  ${\bf x}_{F}\in R^{2}$.
The manifold of matrices is homeomorphic to  $R\times
SL(D-2,R)/O(D-2)$. The corresponding decomposition takes the form
\begin{displaymath}
G=(\det G)^{\frac{1}{D-2}}\left(\left(\det G\right)^{-
\frac{1}{D-2}} G\right) \equiv (\det G)^{\frac{1}{D-2}}
\tilde{G}\equiv (\exp 2\psi)\tilde{G}
\end{displaymath}
We choose a conformal invariant action for $G$
 (this is an  infinite dimensional group  \cite{BPZ}, the
  model is not invariant
under the whole group of diffeomorphims of  the metric, but such  an invariance
is anyway expected to be broken in a quantum theory) \begin{equation}
\begin{array}{l} W(G)=\frac{\alpha}{2}Tr\int d^{2}x_{F}G^{-1}\partial G
G^{-1}\overline{\partial}G + WZW \cr =2\alpha\int
d^{2}x_{F}\partial\psi\overline{\partial}\psi + \frac{\alpha}{2}Tr\int
d^{2}x_{F}\tilde{G}^{-1}\partial\tilde{ G}
\tilde{G}^{-1}\overline{\partial}\tilde{G} + WZW \end{array}
\end{equation} where $\partial=\partial_{1}-i\partial_{2}$ is the
holomorphic derivative and $WZW$ denotes the Wess-Zumino-Witten term
\cite{witten}. It is convenient to choose the following parametrization
for $G$ ($T$ denotes the transposition) \begin{equation} G={\cal
N}\Lambda{\cal N}^{T} \end{equation} where $\Lambda$ is a diagonal
matrix, ${\cal N}$ is a nilpotent matrix with $1$ on the diagonal and
the matrix elements below the diagonal are equal to zero. Then
\begin{equation} e={\cal N}\sqrt{\Lambda} \end{equation} can be chosen
as the tetrad. Using the WZW cocycle condition \begin{displaymath}
W(GH^{-1})= W(G)+W(H^{-1})+2\alpha Tr\int G^{-1}\partial G
H^{-1}\overline{\partial}H \end{displaymath} we can express $W$
explicitly in the coordinates (3) \begin{equation}
W(G)=\frac{\alpha}{2}Tr\int \Lambda^{-1}\partial\Lambda
\Lambda^{-1}\overline{\partial}\Lambda+ \frac{\alpha}{2}Tr\int
(\Lambda^{-1}{\cal N}^{-1} \partial {\cal N})^{T}{\cal N}^{-1}
\overline{\partial}{\cal N}\Lambda \end{equation} This is a model of an
exactly soluble conformal field theory
\cite{haba}\cite{gawedzki}\cite{berlin}. In the following we use the result
that $e={\cal N}\sqrt{\Lambda}$ is a conformal field.
In the coordinates (4) the conformal
fields are $e^{1}_{1}=\exp(\psi+\phi)$, $e^{2}_{2}=\exp(\psi-\phi)$,
 $e^{2}_{1}=\exp(\psi-\phi)X$ and $e^{1}_{2}=0$ where
$\Lambda^{1}_{1}=\exp(2\psi+2\phi)$,$\Lambda^{2}_{2}= \exp
(2\psi-2\phi)$ and the matrix element of ${\cal N}-1$ is denoted by $X$.
The scale dimension of $e$ is $\gamma= -\alpha$ ($\alpha$ must be chosen
negative, this is  inconsistent with a finiteness of the functional
integral but we may work at the begining with a positive $\alpha $ and
only at the level of correlation functions continue $\alpha$ to negative
values; the  relation  of  the functional integral to the free field
representation of conformal field theories  is discussed  in
\cite{gawedzki}). The conformal invariance implies that the random fields
$\lambda^{\gamma}e^{\mu}_{a}(\lambda {\bf x}_{F})$ and $e^{\mu}_{a}(
{\bf x}_{F})$ are equivalent, i.e., they have the same correlation
functions.

\section{The scalar propagator}
We consider a complex scalar matter field $\Phi$ in $D$
dimensions interacting with gravitons depending only on a
$d$-dimensional submanifold. We split the coordinates as
  $x=({\bf x}_{G},{\bf x}_{F})$ with ${\bf x}_{F}\in R^{d}$.
Without a self-interaction the $\Phi \overline{\Phi}$ correlation
function is equal to an average
\begin{equation}
\hbar \int {\cal
D}e\exp\left(\frac{1}{\hbar}W\left(G\right)\right) {\cal
A}^{-1}(x,y)
\end{equation}
over the gravitational field $e$ of the Green's
function of the operator
\begin{equation}
{\cal A}=\frac{1}{2}
\sum_{\mu=1,\nu=1}^{D-d}g^{\mu\nu}({\bf x}_{F})\partial_{\mu}\partial_{\nu}+
\frac{1}{2}\sum_{k=D-d+1}^{D}\partial_{k}^{2}
\end{equation}
We repeat some steps of ref.\cite{PLB} (our case here is
simpler and more explicit). We represent the Green's function by means of the proper time method
\begin{equation}
{\cal A}^{-1}(x,y)=\int_{0}^{\infty}d\tau\left(\exp\left(\tau {\cal A}\right)
\right)(x,y)
\end{equation}
For a calculation of   $\left(\exp\left(\tau {\cal A}\right)
\right)(x,y)   $ we apply the functional integral
\begin{equation}
\begin{array}{l}
K_{\tau}(x,y)=\left(\exp\left(\tau {\cal A}\right)
\right)(x,y)=\int {\cal D}x\exp(-\frac{1}{2}\int \frac{d{\bf x}_{F}}{dt}
  \frac{d{\bf x}_{F}}{dt}-\frac{1}{2}\int g^{\mu\nu}({\bf x}_{F})\frac{dx_{\mu}}{dt}
  \frac{dx_{\nu}}{dt})
 \cr
 \delta\left(x\left(0\right)-x\right)
  \delta\left(x\left(\tau\right)-y\right)
  \end{array}
  \end{equation}
In the functional integral (9) we make a change of variables ($x
\rightarrow b$) determined by Stratonovitch stochastic
differential equations \cite{ike}
\begin{equation}
dx^{\Omega}(s)=e_{A}^{\Omega}\left(
x\left(s\right)\right)db^{A}(s)
\end{equation}
where for $\Omega=1,2,....,D-d$
\begin{displaymath}
e^{\mu}_{a}e^{\nu}_{a}=g^{\mu\nu}
\end{displaymath}
and $e^{\Omega}_{A}=\delta^{\Omega}_{A}$ if $\Omega>D-d$.

As a result of the transformation $x\rightarrow b$
the functional integral becomes Gaussian with the covariance
\begin{displaymath}
E[b_{a}(t)b_{c}(s)]=\delta_{ac}\min(s,t)
\end{displaymath}
In contradistinction to \cite{PLB}
eq.(10) can be solved explicitly. The solution $q_{\tau}$ of eq.(10) consists of two vectors $({\bf
q}_{G},{\bf q}_{F})$ where
\begin{equation}
{\bf q}_{F}(\tau,{\bf x}_{F})={\bf x}_{F}+ {\bf b}_{F}(\tau)
\end{equation}
and ${\bf q}_{G}$ has the components (for $\mu=1,...,D-d$)
\begin{equation}
q^{\mu}(\tau,{\bf x})=x^{\mu}+\int_{0}^{\tau}
e_{a}^{\mu}\left({\bf q}_{F}\left(s,{\bf
x}_{F}\right)\right)db^{a}(s)
\end{equation}
The kernel is
\begin{equation}
\begin{array}{l}
K_{\tau}(x,y)=E[\delta(y-q_{\tau}(x))]= \cr = E[\delta({\bf y}_{F}
-{\bf x}_{F}-{\bf b}_{F}(\tau))
\prod_{\mu}\delta\left(y_{\mu}-q_{\mu} \left(\tau,x\right)\right)]
\end{array}
\end{equation}
Using eq.(12) and the Fourier representation of the $\delta$-function
we write eq.(13) in the form
\begin{equation}
\begin{array}{l}
K_{\tau}(x,y)=(2\pi)^{-D}\int d{\bf p}_{G}d{\bf p}_{F}
\cr
E[\exp\left( i{\bf p}_{F}\left({\bf y}_{F}-{\bf x}_{F}\right)
+i{\bf p}_{G}\left({\bf y}_{G}-{\bf x}_{G}\right)-i{\bf p}_{F}{\bf
b}_{F}\left(\tau\right)-i\int p_{\mu}e^{\mu}_{a}\left
({\bf q}\left(s,{\bf x}_{F}\right)\right)db^{a}\left(s\right)\right)]
\end{array}
\end{equation}
 As discussed in ref.\cite{brze} the expectation value
 in eq.(14) is finite if $\gamma<\frac{1}{2}$.

\section{The scale invariant model}
In general we cannot calculate the average over
the metric explicitly. However, the scale invariance of the
metric is sufficient for a derivation of the
short distance behaviour of  the scalar propagator.

Let us note that     $\sqrt{\tau}b(s/\tau)\simeq \tilde{b}(s)$
where $\tilde{b}$ denotes an equivalent Brownian motion (the
equivalence means that both random variables have the same
correlation functions). Then, using the scale invariance of $e$
with the index $\gamma$ (discussed in sec.2) we can write
\begin{equation}
e(\sqrt{\tau}{\bf x}_{F}) \simeq\tau^{-\frac{\gamma}{2}}\tilde{e}({\bf x}_{F})
\end{equation}
Hence, in eq.(13)
\begin{equation}
q^{\mu}(\tau,{\bf
x})=x^{\mu}+\tau^{\frac{1}{2}-\frac{\gamma}{2}}\int_{0}^{1}
\tilde{e}_{a}^{\mu}\left(\tau^{-\frac{1}{2}} {\bf x}_{F}+
\tilde{{\bf b}}_{F}\left(s\right)\right)d\tilde{b}^{a}(s)
\end{equation}
The expectation value over $e$ is
\begin{equation}
\begin{array}{l}
\langle K_{\tau}(x, y)\rangle= \cr
\tau^{-\frac{D-2}{2}(1-\gamma)-1} \langle E
\left[\delta\left(\left({\bf y}_{F}-{\bf
x}_{F}\right)\tau^{-\frac{1}{2}}-\tilde{{\bf
b}}_{F}\left(1\right)\right)\delta\left(
\tau^{-\frac{1}{2}+\frac{\gamma}{2}}\left(y-x\right)-
\eta\right)\right]\rangle
\end{array}
\end{equation}
where
\begin{equation}
\eta^{\mu}= \int_{0}^{1}
\tilde{e}_{a}^{\mu}\left(\tau^{-\frac{1}{2}}{\bf x}_{F} +
\tilde{{\bf b}}_{F}\left(s\right)\right)d\tilde{b}^{a}(s)
\end{equation}

Let $P({\bf u},{\bf v})$ be the joint distribution of $(\eta,\tilde{{\bf
b}}_{F}(1))$
( $P$ does not depend on ${\bf x}_{F}$ because of the translational
invariance). Then, the propagator of the $\Phi$ field is
\begin{equation}
\begin{array}{l}
\hbar\langle {\cal A}^{-1}( x, y)\rangle= \cr \hbar
\int_{0}^{\infty}d\tau\tau^{-(1-\gamma)\frac{D-2}{2}-1} P\left(
\left({\bf x}_{G}-{\bf y}_{G}\right)\tau^{(-1+\gamma)/2},\left({\bf
x}_{F}-{\bf y}_{F}\right)\tau^{-\frac{1}{2}}\right)
\end{array}
\end{equation}
Eq.(19) in momentum space has the representation
\begin{displaymath}
\hbar\langle {\cal A}^{-1}({\bf k}_{G},{\bf k}_{F})\rangle
=\hbar \int_{0}^{\infty}d\tau \tilde{P}
(\tau^{\frac{1-\gamma}{2}}{\bf k}_{G},\sqrt{\tau}{\bf k}_{F})
\end{displaymath}
where $\tilde{P}$ denotes the Fourier transform of $P$.
Using eq.(14) we may write explicitly
\begin{displaymath}
\begin{array}{l}
\hbar\langle {\cal A}^{-1}({\bf k}_{G},{\bf k}_{F})\rangle
\cr
=\hbar\int_{0}^{\infty}d\tau\langle E[\exp i\left(
\sqrt{\tau}{\bf k}_{F}\tilde{{\bf b}}_{F}\left(1\right)
 +\tau^{\frac{1}{2}-\frac{\gamma}{2}}{\bf k}_{G}\eta_{G}\right)]\rangle
 \end{array}
 \end{displaymath}
The dispersion relation (relating the frequency to the wave number)
 is determined  by  (after an analytic continuation $k_{0}\rightarrow
 ik_{0}$)
\begin{displaymath}
(     \langle {\cal A}^{-1}({\bf k}_{G},{\bf k}_{F})\rangle  )^{-1}=0
\end{displaymath}
It can be concluded from eq.(19) that in general the
dispersion relation will be different from the
standard one (resulting from a non-linear wave equation)
$k_{0}\sim \vert {\bf k}\vert$. In particular, we can see that if $\vert {\bf k}_{F}\vert
\gg \vert {\bf k}_{G}\vert $ then
$\langle {\cal A}^{-1}({\bf k}_{G},{\bf k}_{F})\rangle\sim
\vert {\bf k}_{F}\vert^{-2}$ whereas if    $\vert {\bf k}_{G}\vert
\gg \vert {\bf k}_{F}\vert $     then
$ \langle {\cal A}^{-1}({\bf k}_{G},{\bf k}_{F})\rangle \sim
 \vert {\bf k}_{G}  \vert^{-\frac{2}{1-\gamma}}$.
 In the configuration space, the propagator tends to infinity if both $\vert
{\bf x}_{F}- {\bf y}_{F}\vert$ and $\vert {\bf x}_{G}- {\bf
y}_{G}\vert$  tend to zero. However, the singularity depends in a
rather complicated way on the approach to zero. It becomes simple
if either  $\vert {\bf x}_{F}- {\bf y}_{F}\vert=0$ or $\vert {\bf
x}_{G}- {\bf y}_{G}\vert=0$   . So, if    $\vert {\bf x}_{F}- {\bf
y}_{F}\vert=0$ then we make a change of the time variable
\begin{equation}
\tau=t\vert {\bf x}_{G}-{\bf y}_{G}\vert ^{\frac{2}{1-\gamma}}
\end{equation}
Using eq.(19) we obtain the factor   depending on $\vert {\bf
x}_{G}-{\bf y}_{G}\vert    $ in front of the integral and a
bounded function $A$ of coordinates ,i.e.,
\begin{equation}
\langle {\cal A}^{-1}( x, y)\rangle= A\vert {\bf x}_{G}-{\bf
y}_{G}\vert^{-D+2}
\end{equation}
If            $\vert {\bf x}_{G}-
{\bf y}_{G}\vert=0$
then we change the time variable
\begin{displaymath}
\tau=t  \vert {\bf x}_{F}-
{\bf y}_{F}\vert^{2}
\end{displaymath}
As a result
\begin{equation}
\langle {\cal A}^{-1}( x, y)\rangle= A \vert {\bf x}_{F}- {\bf
y}_{F}\vert^{-(D-2)(1-\gamma) }
\end{equation}
with a certain bounded function $A$.
We can see that in the ${\bf x}_{G}$ coordinate the singularity
remains unchanged but the propagator is more regular
in the ${\bf x}_{F}$ coordinate.

It is not easy to calculate the probability distribution $P$
exactly. Choosing as a first approximation $\eta\simeq {\bf
b}_{G}(1)$  we obtain \begin{displaymath} P({\bf u},{\bf
v})=(2\pi)^{-\frac{D}{2}}\exp(-\frac{{\bf u}^{2}}{2} -\frac{{\bf
v}^{2}}{2})
\end{displaymath}
In this approximation
\begin{displaymath}
\hbar\langle {\cal A}^{-1}({\bf k}_{G},{\bf k}_{F})\rangle
=\frac{\hbar}{2} \int_{0}^{\infty}d\tau \exp(- \frac{1}{2}
\tau^{1-\gamma}\vert {\bf k}_{G}\vert^{2}-\frac{1}{2}\tau\vert {\bf k}_{F}\vert^{2})
\end{displaymath}
If $D=4$ then
\begin{equation}
\int d{\bf x}_{F}d{\bf x}_{G}
\vert\langle {\cal A}^{-1}( x, y)\rangle\vert ^{2}=
\int d{\bf k}_{G}d{\bf k}_{F}\vert\langle { \cal A}^{-1}(k)\rangle\vert^{2}  <\infty
\end{equation}
for any $\gamma>0$.

In $D$ dimensions the integral (23) takes the form
\begin{displaymath}
\int_{0}^{\infty}d\tau_{1}\int_{0}^{\infty}d\tau_{2}
(\tau_{1}+\tau_{2})^{-1}(\tau_{1}^{1-\gamma}
+\tau_{2}^{1-\gamma})^{-\frac{D-2}{2}}
\end{displaymath}
It is finite if $(1-\gamma)(D-2)<2$. Hence, it can be finite
in $D=5$ if $\gamma<\frac{1}{2}$ is large enough. In
$D=6$ power-law singularities appear if $\gamma<\frac{1}{2}$
and the model has logarithmic (presumably renormalizable)
singularities for $\gamma=\frac{1}{2}$.

 It can be shown that
a singularity of the mean value of any power $n$
of ${\cal A}^{-1}( x, y)$ is equal to the $n$-th power of the
singularity of the mean value  $\langle {\cal A}^{-1}\rangle$. Then, it follows that there will be no
coupling constant renormalization in the $\Phi^{4}$
model in four dimensions.
\section{The photon propagator}
 Let us consider the $D$-dimensional electromagnetic Lagrangian
\begin{equation}
W =\frac{1}{4}\int d^{D}x\sqrt{g}g^{BR}g^{CS}F_{BC}F_{RS}
\end{equation}
We define
\begin{equation}
\hat{g}^{BC}=\sqrt{g}g^{BC}
\end{equation}
and
\begin{equation}
\hat{g}^{BC}=\hat{e}^{B}_{S}\hat{e}^{C}_{S}
\end{equation}
We consider also the inverse matrix $l$
\begin{equation}
\hat{l}_{B}^{S}\hat{e}^{B}_{R}=\delta^{S}_{R}
\end{equation}
In terms of the tetrad we define fields in a local Euclidean frame
\begin{equation}
\hat{A}_{R}=\hat{e}^{B}_{R}A_{B}
\end{equation}
We choose the Feynman gauge. This means an addition to the action
of the term
\begin{equation}
W_{0}=\frac{1}{2}\int d^{D}x(g^{BC}\partial_{B}\sqrt{g}A_{C})^{2}
\end{equation}
We express the action by the hat-fields. We divide
$W=W_{1}+W_{2}+W_{3}+W_{0}$ as follows
\begin{equation}
W_{1} =\frac{1}{2}\int d^{D}x\hat{A}_{a}(-g^{\rho\sigma}
\partial_{\rho}\partial_{\sigma})\hat{A}_{a}
\end{equation}
where $ \hat{A}_{a}=\hat{e}_{a}^{\nu}A_{\nu}$
and
 \begin{equation}
W_{2} =\frac{1}{2}\int d^{D}x\hat{A}_{k}(
(-\partial_{j}+\hat{\Gamma}_{j})(\partial_{j}+\hat{\Gamma}_{j})
-g^{\rho\sigma}
\partial_{\rho}\partial_{\sigma})\hat{A}_{k}
\end{equation}
where
\begin{equation}
(\hat{A})_{k}=g^{\frac{1}{4}}A_{k}
\end{equation}
is a vector and
\begin{equation}
(\hat{\Gamma}_{j})^{l}_{k}=g^{\frac{1}{4}}\partial_{j}
g^{-\frac{1}{4}}\delta_{k}^{l}
\end{equation}
is a diagonal matrix .
$\hat{\Gamma}$ is a connection defined in eq. (33) for the last
$(D-1,D)$ components and
\begin{equation}
\hat{\Gamma}^{c}_{ja}=\hat{e}^{\mu}_{a}\partial_{j}\hat{l}^{c}_{\mu}
\end{equation}
for the first $D-2$ components. After the transformation (28)
$W_{3}$ reads
\begin{equation}
\begin{array}{l}
W_{3}=\frac{1}{2}\int (\partial_{j}\delta_{ca}+\hat{\Gamma}^{a}_{jc})\hat{A}_{a}
 (\partial_{j}\delta_{cb}+\hat{\Gamma}^{b}_{jc})\hat{A}_{b}
\cr
\equiv \frac{1}{2}\int \hat{A}_{a} [(-\partial_{j}+\hat{\Gamma}_{j})
 (\partial_{j}+\hat{\Gamma}_{j})]_{ab}\hat{A}_{b}
 \end{array}
 \end{equation}
 We can write down the whole action  in the form
 \begin{equation}
 W=\frac{1}{2}\int d^{D}x \hat{A}_{R} [g^{BC}(-\partial_{B}+\hat{\Gamma}_{B})
 (\partial_{C}+\hat{\Gamma}_{C})]_{RS}\hat{A}_{S}
\equiv \frac{1}{2}\int \hat{A}_{R}(-\triangle_{EM})_{RS}\hat{A}_{S}
 \end{equation}
 where $\hat{\Gamma}_{j}$ has a block form with components defined in
 eqs.(33)
 and (34) ($\hat{\Gamma}_{\mu}=0$).

For a calculation of the correlation functions in QFT expressed by
$\triangle_{EM}^{-1}$ we are interested in the heat equation
\begin{equation}
\partial_{\tau}\hat{A}=\frac{1}{2}\triangle_{EM}\hat{A}
\end{equation}
The solution can be expressed in a general form \cite{ike}
\begin{equation}
\hat{A}(\tau,x)\equiv
(T_{\tau}\hat{A})(x)=E[{\cal T}_{\tau}\hat{A}(q_{\tau}(x))]
\end{equation}
where $\hat{A}_{B}$ is a $D$-dimensional vector and the matrix
${\cal T}$ is a solution of the equation
\begin{equation}
d {\cal T}=\hat{\Gamma}_{j}{\cal T} db^{j}
\end{equation}
where $b^{j}$ are independent Brownian motions.

We need the kernel $K$ of the operator $T_{\tau}$ defined
as the solution of the equation
\begin{equation}
\partial_{\tau}T_{\tau}=\frac{1}{2}\triangle_{EM}T_{\tau}
\end{equation}
with the initial condition $(T_{\tau})_{\vert \tau=0}=1$.
In components, eq.(40)  reads
\begin{equation}
\partial_{\tau}K_{R}^{S}(\tau;x,y)=\frac{1}{2}
\int d^{D}z\triangle_{EM}(x,z)_{R}^{C}K(\tau;z,y)_{C}^{S}
\end{equation}
with the initial condition $ K(\tau=0;z,y)_{D}^{S}=\delta_{D}^{S}
\delta(z-y)$   .     Note that
\begin{equation}
\hat{\Gamma}_{j}=\hat{e}\partial_{j}\hat{e}^{-1}
\end{equation}
 is of the form of a pure gauge. Then, eq.(39) can be solved
 exactly.
Inserting the solution of eq.(39) into eq.(38) we obtain the
solution of the heat equation in the form
\begin{equation}
\hat{A}_{R}(\tau,x)=E[\hat{e}^{C}_{R}(x)\hat{l}_{C}^{S}(q_{\tau}(x))
\hat{A}_{S}(q_{\tau}(x))]
\end{equation}
Hence, the kernel is
\begin{equation}
K_{R}^{S}(\tau;x,y)=\hat{e}_{R}^{D}(x)\hat{l}_{D}^{S}(y)
E[\delta(y-q_{\tau}(x))   ]
\end{equation}
We must take an average over the tetrad in order to compute
correlations in a random metric field. It is not easy to
calculate expectation values of the inverse of $\hat{e}$. So, we
restrict ourselves to some special correlations .

The formula
\begin{equation}
\langle \hat{A}_{R}(x)\hat{A}_{P}(y)\rangle=\langle
(-\triangle_{EM})^{-1}_{RP}(x,y)\rangle \equiv \langle {\cal
D}_{RP}(x,y)\rangle
\end{equation}
follows from the action (36). We can use this result to compute
\begin{equation}
\begin{array}{l}
\langle {\cal D}_{B}^{C}(x,y)\rangle\equiv\langle
A_{B}(x)A^{C}(y)\rangle=\langle A_{B}(x)g^{CD}(y)A_{D}(y)\rangle
\cr =\langle\hat{l}_{B}^{R}(x)\hat{A}_{R}(x)\hat{e}^{C}_{S}(y)
\hat{e}^{D}_{S}(y) \hat{l}_{D}^{P}(y)\hat{A}_{P}(y) \rangle
 = \int_{0}^{\infty} d\tau\langle\hat{l}_{B}^{R}(x)\hat{e}^{C}_{S}(y) \hat{e}^{D}_{S}(y)
\hat{l}_{D}^{P}(y)K^{P}_{R}(\tau;x,y) \rangle
\cr
=\delta_{B}^{C}\int_{0}^{\infty}d\tau \langle
E[\delta(y-q_{\tau}(x))]\rangle
 \end{array}
 \end{equation}
 where the r.h.s. is equal to the scalar propagator.
 We can conclude that the short distance behaviour
 of the scalar as well as photon propagators are the same.
 It is less singular than the canonical one and determined
 by the scale dimension of $e$.

Finally, we consider  the Abelian Higgs model.
Its gauge invariant perturbation expansion in the  $(\Phi\Phi^{*})^{2}$
interaction is expressed by the propagators
\begin{equation}
\begin{array}{l}
\langle \Phi(x)\Phi^{*}(y)\rangle=
\langle\int_{0}^{\infty}d\tau E[\delta\left(q\left(\tau,x\right)-y\right)
\exp\left(i\int_{0}^{\tau}dq^{B}(s)A_{B}\left (q\left(s\right)\right)\right)
]\rangle
 \cr
=\langle\int_{0}^{\infty}d\tau
E[\delta\left(q\left(\tau,x\right)-y\right)
\exp\Big(-\frac{1}{2}\int_{0}^{\tau}dq^{B}(s)dq_{C}(s^{\prime}){\cal
D}_{B}^{C} \left (q\left(s\right)-q\left(s^{\prime}\right)\right)
\Big)]\rangle
\end{array}
\end{equation}
where ${\cal D}$ has been calculated in eq.(46). In the
approximation
 $\langle \exp(-{\cal D})\rangle\simeq \exp(-\langle {\cal D}\rangle)$ we
 obtain
\begin{equation}
\begin{array}{l}
\langle \Phi(x)\Phi^{*}(y)\rangle\cr
\simeq\int_{0}^{\infty}d\tau E[\langle\delta\left(q\left(\tau,x\right)-y\right)
\rangle\exp\Big(-\frac{1}{2}\langle\int_{0}^{\tau}dq^{B}(s)dq_{C}(s^{\prime})
 {\cal D}_{B}^{C}
\left(q\left(s\right)-q\left(s^{\prime}\right)\right)\rangle\Big)
]
\end{array}
\end{equation}
 The double line integral in the exponential (48) is more
regular than the canonical one and may be finite depending on the
dimension $D$ and the value of $\gamma$. In particular, it is
finite if $D=4$ . The expression for higher order correlation
functions is similar to eq.(47). So, for the fourth order
correlation function we have
\begin{equation}
\begin{array}{l}
\langle \Phi(x)\Phi(x^{\prime})\Phi^{*}(y)\Phi^{*}(y^{\prime})\rangle
=\langle\int_{0}^{\infty}d\tau d\tau^{\prime}
 E[\delta\left(q\left(\tau,x\right)-y\right)
\delta\left(q^{\prime}\left(\tau^{\prime},x^{\prime}\right)-y^{\prime}\right)
\cr
\exp\left(i\int_{0}^{\tau}dq_{B}A^{B}+i\int_{0}^{\tau^{\prime}}
dq_{B}^{\prime}A^{B}\right)
]\rangle     +(x\rightarrow x^{\prime})
 \cr
=\langle\int_{0}^{\infty}d\tau d\tau^{\prime}
E[\delta\left(q\left(\tau,x\right)-y\right)
\delta\left(q^{\prime}\left(\tau^{\prime},x^{\prime}\right)-y^{\prime}\right)
\cr
\exp\Big(-\frac{1}{2}\int_{0}^{\tau}\int_{0}^{\tau}dq^{B}(s)dq_{C}(s^{\prime}){\cal
D}_{B}^{C} \cr
-\frac{1}{2}\int_{0}^{\tau^{\prime}}\int_{0}^{\tau^{\prime}}
dq^{B\prime}(s)dq_{C}^{\prime}(s^{\prime}){\cal D}_{B}^{C}
-\int_{0}^{\tau^{\prime}}\int_{0}^{\tau}dq^{B}(s)dq_{C}^{\prime}(s^{\prime}){\cal
D}_{B}^{C} \Big) ]\rangle \cr +(x\rightarrow x^{\prime})
\end{array}
\end{equation}
where the last term means the same expression with $x$ replaced
by $x^{\prime}$. We can obtain similar formulae for higher order
correlation functions (including the correlations with the gauge
potential). It follows that as a result of the more regular short
distance behaviour of the scalar propagator (19) and the
electromagnetic propagator (45) there will be no charge
renormalization in the fourdimensional Higgs field interacting
with a scale invariant two-dimensional quantum gravity. In higher
dimensions $D$ the charge renormalization depends on the values
of $D$ and  $\gamma$. There still will be no charge
renormalization if $D=5$ and $\gamma $ is sufficiently close to
$\frac{1}{2}$.

\end{document}